\documentclass[a4paper,11pt]{article}

\usepackage{eucal}
\usepackage{amsfonts,amsmath}
\usepackage{graphics,epsfig}
\usepackage{mathrsfs,latexsym}
\usepackage{fancyhdr}
\usepackage{cite}

\def\bbc{{\Bbb C}}
\def\bbr{{\Bbb R}}

\def\bbz{{\Bbb Z}}

\def\tr{\mathrm{tr\,}}

\def\re{\mathrm{Re\,}}
\def\im{\mathrm{Im\,}}
\def\ad{\mathrm{ad\,}}

\def\mod{\text{mod\,}}

\def\rmd{\mathrm{d\,}}
\def\rmi{\mathrm{i}}
\def\rme{\mathrm{e}}

\def\openone{\leavevmode\hbox{\small1\kern-3.3pt\normalsize1}}
\def\diag{\mbox{diag\,}}

\newtheorem{rem}{Remark}

\begin{document}
\title{Nonlocal Reductions of a Generalized Heisenberg Ferromagnet Equation}
\author{R. Myrzakulov$^1$, G. Nugmanova$^1$, T. Valchev$^2$ and K. Yesmakhanova$^1$\\
\small $^1$ Department of General \& Theoretical Physics,\\
\small Eurasian National University, Astana, 010008, Kazakhstan\\
\small $^2$ Institute of Mathematics and Informatics,\\
\small Bulgarian Academy of Sciences, Acad. G. Bonchev Str.,\\
\small 1113 Sofia, Bulgaria\\
\small E-mails: tiv@math.bas.bg, rmyrzakulov@gmail.com,\\
\small nugmanovagn@gmail.com, kryesmakhanova@gmail.com}
\date{}
\maketitle
\begin{abstract}
We study nonlocal reductions of coupled equations in $1+1$ dimensions of the Heisenberg ferromagnet
type. The equations under consideration are completely integrable and have a Lax pair related to a linear
bundle in pole gauge. We describe the integrable hierarchy of nonlinear equations related to our system
in terms of generating operators. We present some special solutions associated with four distinct discrete
eigenvalues of the scattering operator. Using Lax pair diagonalization method, we derive recurrence formulas
for the conserved densities and find the first two simplest conserved densities. 
\end{abstract}

\tableofcontents

\newpage
\section{Introduction}\label{intro}

Heisenberg ferromagnet equation (HF)
\begin{equation}
\mathbf{S}_t = \mathbf{S}\times\mathbf{S}_{xx} 
\label{hf}\end{equation}
for the three dimensional vector $\mathbf{S}(x,t)$ of unit length is one of classical soliton equations \cite{BoPo90, book,blue-bible}.
Above, "$\times$" stands for the standard cross product in the three dimensional Euclidean space and the subscripts denote partial derivatives with respect to the variables $t$ and $x$. The vector equation (\ref{hf}) has been subject to various generalizations,
e.g. the following coupled system of nonlinear evolution equations (NEEs)
\begin{equation}
\begin{split}
\rmi u_t & + u_{xx}+[(\varepsilon u\overline{u}_x +  v\overline{v}_x)u]_x =  0\, , \qquad \varepsilon = \pm 1 \, ,\\
\rmi v_t & + v_{xx}+[(\varepsilon u\overline{u}_x + v\overline{v}_x)v]_x = 0\, ,
\end{split}	\label{ghf}	
\end{equation}
was proposed as an integrable generalization of HF, see \cite{side9,yan,yantih,yantih2,yanvil}.
Above, overlining means complex conjugation and "$\rmi$" is the imaginary unit. The dynamical fields $u$ and $v$
are required to satisfy the additional condition
\begin{equation}
\varepsilon \vert u\vert^2 + \vert v\vert^2 = 1 
\label{constr_0}\end{equation}
where $\vert . \vert$ is the usual norm in the complex plane ($\vert z\vert = \sqrt{z\overline{z}}$, $z\in\bbc$). The system
(\ref{ghf}) is the compatibility condition $[L(\lambda),A(\lambda)] = 0$ of the Lax operators:
\begin{eqnarray}
L(\lambda) & = & \rmi\partial_x - \lambda S, \qquad \lambda\in\bbc,\qquad
S = \left(\begin{array}{ccc}
0 & u & v\\ 
\varepsilon \overline{u} & 0 & 0 \\
\overline{v} & 0 & 0\end{array}\right), \label{lax1}\\
A(\lambda) &=& \rmi\partial_t + \lambda A_1 + \lambda^2 A_2,
\qquad A_2  = \left(\begin{array}{ccc}
- 1/3 & 0 & 0 \\ 0 & 2/3 - \varepsilon \vert u\vert^2 & -\varepsilon \overline{u}v \\
0 & - \overline{v} u & 2/3 - \vert v\vert^2
\end{array}\right),\label{lax2}\\
A_1 & = & \left(\begin{array}{ccc}
0 & a & b\\ 
\varepsilon \overline{a} & 0 & 0 \\
\overline{b} & 0 & 0\end{array}\right),\quad
\begin{array}{ccc} 
a & = & - \rmi u_x - \rmi\left(\varepsilon u\overline{u}_x + v\overline{v}_x\right)u\\
b & = & - \rmi v_x - \rmi\left(\varepsilon u\overline{u}_x + v\overline{v}_x\right)v
\end{array}
\label{lax3}
\end{eqnarray}
which generalizes in a natural way the Lax pair of (\ref{hf}), see \cite{blue-bible}.

Another possible way to generalize (\ref{hf}) is by constructing multidimensional NEEs. Examples
of $2+1$ dimensional generalizations of HF are given by Ishimori's equation \cite{ishim}
\begin{eqnarray*}
\mathbf{S}_t = \mathbf{S}\times (\mathbf{S}_{xx} + \mathbf{S}_{yy}) + u_x\mathbf{S}_y + u_y \mathbf{S}_x\\
u_{xx} - u_{yy} =  - 2\mathbf{S} . (\mathbf{S}_x \times \mathbf{S}_y)
\end{eqnarray*}
and Myrzakulov I equation \cite{myr1,myr2}
\[\mathbf{S}_t = \mathbf{S}\times \mathbf{S}_{xy} + u\mathbf{S}_x, \qquad
u_{x} =  - \mathbf{S} . (\mathbf{S}_x \times \mathbf{S}_y).\]
In both equations above, $\mathbf{S}$ is a three dimensional vector of unit length, $u$ is a real valued function
and "dot" denotes the usual scalar product of vectors in the three dimensional Euclidean space.

A recent trend in the theory of integrable systems, initiated by Ablowitz and Musslimani \cite{ablomus}, is the study
of nonlocal reductions of NEEs. Nowadays, finding nonlocal counterparts corresponding to well-known (local) NEEs
enjoys an ever increasing interest, see \cite{fok,ggi,gurpec} and references therein. Our main purpose here is to study
a nonlocal counterpart of (\ref{ghf}), namely the following system of NEEs
\begin{equation}
\begin{split}
\rmi u_t(x,t) + u_{xx}(x,t) + \left\{\left[\varepsilon u(x,t)\overline{u_x(-x,t)}
+ v(x,t)\overline{v_x(-x,t)}\right]u(x,t)\right\}_x & = 0, \\
\rmi v_t(x,t) + v_{xx}(x,t) + \left\{\left[\varepsilon u(x,t)\overline{u_x(-x,t)}
+ v(x,t)\overline{v_x(-x,t)}\right]v(x,t)\right\}_x & = 0
\end{split}	
\label{ghf_a}\end{equation} 
for the dynamical fields $u$ and $v$. That coupled system is solvable through inverse scattering transform,
i.e. it has a Lax representation, soliton type solutions, an infinite number of conservation laws etc.

The text of the preprint is organized in the following manner. In next section, we shall introduce the Lax pair of
(\ref{ghf_a}), then make some remarks on the structure of the Lax operators and their symmetries (reductions). The
third section is dedicated to the integrable hierarchy of NEEs related to the coupled system under consideration.
Starting from a general flow Lax pair, we shall show how the hierarchy can be described in terms of recursion operators.
Section \ref{solutions} contains discussion on how we can integrate (\ref{ghf_a}) by using dressing method
\cite{zakh-mikh,ZS} (Darboux transformation method). Following the algorithm described in \cite{yantih2} for the case
of linear bundles in pole gauge, allows us to construct a simple class of special solutions over constant background.
In Section \ref{consdens}, we describe an algorithm to derive the integrals of motion of (\ref{ghf_a}). For that
purpose we shall apply Lax pair diagonalization method \cite{DrSok*85} that will allow us to find a recursive formula
for the corresponding conserved densities. Section "Conclusion" contains our final remarks and some further discussion
on our results.

\section{Lax Representation}\label{laxpair}

As we mentioned, the system of nonlocal equations (\ref{ghf_a}) is integrable through inverse scattering
transform. Here we shall pay certain attention to its Lax representation. This is the reason why we shall
briefly remind the reader the notion of (nonlocal) reduction thus following \cite{mikh2,tih}. 

Let us introduce the Lax operators
\begin{eqnarray}
L(\lambda) & = & \rmi\partial_x - \lambda S(x,t), \qquad \lambda\in\bbc,\quad
\label{lax_gen1}\\
A(\lambda) &=& \rmi\partial_t + \sum_{j=1}^{N}\lambda^jA_j(x,t),\qquad N\geq 2\label{lax_gen2}
\end{eqnarray}
where all the coefficients $S(x,t)$ and $A_j(x,t)$, $j=1,2,\ldots, N$ are some complex traceless
$3\times 3$ matrices. Let us denote by $\mathscr{F}$ the space of all the fundamental sets of solutions
to the auxiliary linear problem
\begin{equation}
\rmi\partial_x\Psi(x,t,\lambda) - \lambda S(x,t)\Psi(x,t,\lambda) = 0 .
\label{sp_problem}\end{equation}
Since $S(x,t)$ has a zero trace, any fundamental solution $\Psi(x,t,\lambda)$ is unimodular, i.e. we have
$\det\Psi(x,t,\lambda) = 1$. Assume now a finite group $G_{\rm R}$ acts on $\mathscr{F}$ as follows:
\begin{equation}
\mathcal{K}_g : \Psi(x,t,\lambda)\to\tilde{\Psi}(x,t,\lambda) =
\mathrm{K}_g\left[\Psi\left(\kappa^{-1}_{g}(x,t),k^{-1}_g(\lambda)\right)\right],
\qquad g\in\mathrm{G}_{\mathrm{R}}
\label{gr_psi}\end{equation}
where $\kappa_g:\bbr^2\to\bbr^2$ is a smooth, invertible mapping; $k_g:\bbc\to\bbc$ is a conformal 
mapping and $\mathrm{K}_g$ is a group automorphism of the Lie group $\mathrm{SL}(3,\bbc)$. $G_{\rm R}$
transforms the Lax operators according to
\begin{eqnarray}
L(\lambda) & \to & \tilde{L}(\lambda) = \mathcal{K}_g\circ L(\lambda)\circ\mathcal{K}^{-1}_g, 
\label{l_trans}\\
A(\lambda) & \to & \tilde{A}(\lambda) = \mathcal{K}_g\circ A(\lambda)\circ\mathcal{K}^{-1}_g.
\end{eqnarray}
In view of the equality
\[[\tilde{L}(\lambda), \tilde{A}(\lambda)]	=
\mathcal{K}_g\circ[L(\lambda), A(\lambda)] \circ\mathcal{K}^{-1}_g \]
the transformed operators $\tilde{L}(\lambda)$ and $\tilde{A}(\lambda)$ still commute.

The requirement that (\ref{sp_problem}) is $G_{\rm R}$-invariant, implies that
\[\tilde{L}(\lambda) \propto L(\lambda)\]
which imposes certain symmetry condition on the matrix coefficient $S(x,t)$. Effectively, such
condition decreases (reduces) the number of independent entries of $S(x,t)$, i.e. the dynamical fields.
Similar argument holds for the coefficients of the operator $A(\lambda)$ since those can be
expressed through $S$ and its $x$-derivatives. Due to all this, $G_{\rm R}$ is called reduction group
while (\ref{gr_psi}) or equivalently the afore-mentioned symmetries of $S(x,t)$ and $A_j(x,t)$, are called
nonlocal reduction.

\begin{rem}
The term "nonlocal" refers to the fact that $G_{\rm R}$ acts on the independent variables so the zero 
curvature condition leads to a NEE having nonlocal terms. In the particular case when
$\kappa_g = \rm{id}_{\bbr^2}$, $\forall g\in G_{\rm R}$ reduction is called local.
\end{rem}

In order to illustrate the above ideas, let us consider the following reduction
\begin{equation}
\Psi(x,t,\lambda)\to\tilde{\Psi}(x,t,\lambda) = H\Psi(x,t,-\lambda)H,\qquad H=\diag(-1,1,1).
\label{psired1}
\end{equation}
So we require that for any $\Psi\in\mathscr{F}$, $\tilde{\Psi}$, as given in (\ref{psired1}), is another
fundamental solution to (\ref{sp_problem}). Since (\ref{psired1}) involves involutions only ($H^2 = \openone$,
$\openone$ is the identity matrix), it defines an action of the group $\bbz_2$. From  (\ref{psired1}) one
immediately gets
\begin{equation}
HS(x,t)H = - S(x,t) ,\qquad HA_{j}(x,t)H = (-1)^{j}A_{j}(x,t),\qquad j = 1,2 ,\ldots, N .\label{red1}
\end{equation}
Taking into account the explicit form of $H$, we deduce from (\ref{red1}) that the matrix coefficients
of the Lax pair must have the following block structure:
\[S=\left(\begin{array}{ccc}
0 & \ast & \ast \\ \ast & 0 & 0 \\
\ast & 0 & 0 
\end{array}\right),\qquad A_{2k-1} = \left(\begin{array}{ccc}
0 & \ast & \ast \\ \ast & 0 & 0 \\
\ast & 0 & 0 
\end{array}\right),\qquad A_{2k} = \left(\begin{array}{ccc}
\ast & 0 & 0 \\ 0 & \ast & \ast \\
0 & \ast & \ast 
\end{array}\right).\] 

Let us now impose another $\bbz_2$-reduction of the form:
\begin{equation}
\Psi(x,t,\lambda) \to \tilde{\Psi}(x,t,\lambda) = \mathcal{E}\left[\Psi^{\dag}(-x,t,-\overline{\lambda})\right]^{-1}\mathcal{E}\label{psired2}
\end{equation}
where $\mathcal{E} = \diag(1,\varepsilon,1)$, $\varepsilon^2 = 1$ and the symbol "$\dag$" is for Hermitian conjugation.
In that case we have the following symmetry conditions:
\begin{eqnarray}
\mathcal{E}S^{\dag}(-x,t)\mathcal{E}  &=& S(x,t) ,\qquad \mathcal{E} A_{j}^{\dag}(-x,t)\mathcal{E}
= (-1)^{j}A_{j}(x,t). \label{red2}
\end{eqnarray}
The reductions (\ref{psired1}) and (\ref{psired2}) commute so they both can be viewed as an action of the group
$\bbz_2\times\bbz_2$. After taking into account (\ref{psired1}) and (\ref{psired2}), the matrix $S(x,t)$ can be
written down as:
\begin{equation}
S(x,t) = \left(\begin{array}{ccc}
0 & u(x,t) & v(x,t)\\ 
\varepsilon \overline{u(-x,t)} & 0 & 0 \\
\overline{v(-x,t)} & 0 & 0\end{array}\right)
\label{lax1a}\end{equation}
for some complex valued functions $u$ and $v$. From that point on we shall assume that $u$ and $v$ are
not entirely independent functions but satisfy the following nonlocal constraint:
\begin{equation}
\varepsilon u(x,t)\overline{u(-x,t)} + v(x,t)\overline{v(-x,t)} = 1.
\label{constr}
\end{equation}
Such constraint is essential when one tries to employ inverse scattering transform to the Lax pair under consideration,
see \cite{side9} for more explanations.

\begin{rem}
We note that in view of (\ref{constr}) the matrix-valued function $S$ satisfies:
\begin{equation}
S^3 = S.
%\label{s_rel}
\label{s_spectr}\end{equation}
Hence $S$ can be put into diagonal form and its eigenvalues are $-1$, $0$ and $1$.
\label{rem_constr}\end{rem}

Let's now restrict ourselves with the simplest nontrivial case of quadratic flow Lax pair
\begin{equation}
A(\lambda) = \rmi\partial_t + \lambda A_1(x,t) + \lambda^2 A_2(x,t).\label{lax2a}
\end{equation}
Taking into account (\ref{lax1a}) and (\ref{constr}), it is not hard to check that the compatibility
condition $[L(\lambda),A(\lambda)] = 0$ of (\ref{lax_gen1}) and (\ref{lax2a}) determines the
matrix coefficients of (\ref{lax2a}) to be\footnote{Compare these expressions with those in the local 
case, see (\ref{lax1}), (\ref{lax2}) and (\ref{lax3}).}:
\begin{eqnarray}
A_1(x,t) &=& \left(\begin{array}{ccc}
0 & a(x,t) & b(x,t)\\ 
-\varepsilon \overline{a(-x,t)} & 0 & 0 \\
-\overline{b(-x,t)} & 0 & 0\end{array}\right),\nonumber\\
a(x,t) &=& - \rmi u_x(x,t) - \rmi \mathcal{B}(x,t)u(x,t),\qquad b(x,t) =  - \rmi v_x(x,t) - \rmi \mathcal{B}(x,t)v(x,t),
\nonumber\\
\mathcal{B}(x,t) &=& \varepsilon u(x,t)\overline{u_x(-x,t)} + v(x,t)\overline{v_x(-x,t)},\label{cal_b}\\
A_2(x,t) &=& \left(\begin{array}{ccc}
-1/3 & 0 & 0 \\ 0 & 2/3 - \varepsilon \overline{u(-x,t)}u(x,t) & - \varepsilon \overline{u(-x,t)}v(x,t)\\
0 & - \overline{v(-x,t)}u(x,t) & 2/3 - \overline{v(-x,t)}v(x,t)
\end{array}\right).\nonumber
\end{eqnarray}
Moreover, $u$ and $v$ must solve the following coupled system of NEEs:
\begin{eqnarray}
\rmi u_t(x,t) + u_{xx}(x,t) + [\mathcal{B}(x,t)u(x,t)]_x &=& 0, \label{ghf_nonloc1}\\
\rmi v_t(x,t) + v_{xx}(x,t) + [\mathcal{B}(x,t)v(x,t)]_x &=& 0. \label{ghf_nonloc2}
\end{eqnarray}
That nonlocal system represents the main object of interest in the current preprint. As it is clear from the 
derivation itself, (\ref{ghf_nonloc1}) and (\ref{ghf_nonloc2}) represent the simplest member of an
infinite family of nonlocal NEEs each corresponding to an operator $A(\lambda)$ of particular
degree\footnote{The reader may notice that we ignored here the case when $\deg(A(\lambda)) = 1$.
We did it intentionally because the linear flow yields to the following system of linear wave equations
\[u_t + c u_x = 0,\qquad v_t+ c v_x = 0, \qquad c\in\bbr\]
that is not interesting in the context of inverse scattering transform and its applications.}.
That family (hierarchy) will be described in the section to follow.

\section{Integrable Hierarchies}\label{sec_hier}

This section is dedicated to a description of the hierarchy of integrable NEEs associated with coupled system
(\ref{ghf_nonloc1}) and (\ref{ghf_nonloc2}). Below we shall follow ideas and methods discussed in more detail in
\cite{book,side9}. 

Let us consider again the general Lax operators (\ref{lax_gen1}) and (\ref{lax_gen2}) and require the conditions
(\ref{red1}) and (\ref{red2}) hold true. Then $S(x,t)$ is given by the matrix (\ref{lax1a}). We assume that the
constraint (\ref{constr}) is satisfied by the dynamical fields $u$ and $v$.

Let us analyze the compatibility condition $[L(\lambda), A(\lambda)] = 0$ of (\ref{lax_gen1}) and (\ref{lax_gen2}).
After comparing the coefficients before the same powers of $\lambda$, we derive the following string of differential
relations:
\begin{eqnarray}
\lambda^{N+1} &:&	[S, A_N] = 0 , \label{rec_np1}\\
\lambda^N &:& \rmi\partial_xA_N - [S,A_{N-1}] = 0,\label{rec_n}\\
\lambda^j &:& \rmi\partial_xA_j - [S,A_{j-1}] = 0,
\qquad j=2,\ldots, N-1 ,\label{rec_k}\\
\lambda &:& \partial_xA_1 + \partial_t S = 0\label{rec_1}
\end{eqnarray}
for the coefficients of the second Lax operator $A(\lambda)$. For any integer $N\geq 2$, these recurrence relations
yield to some NEE. This way we have a whole family of NEEs, called integrable hierarchy. In order to resolve (\ref{rec_np1})--(\ref{rec_1}), we start from the highest degree equality. Equation (\ref{rec_np1})
tells us that for the highest degree coefficient we have:
\begin{equation}
A_{N} = \left\{\begin{array}{rl} 
c_NS ,& N \equiv 1 \quad (\mod 2) \\ c_N S_1 ,& N \equiv 0 \quad (\mod 2)
\end{array}\right. ,
\label{A_N}\end{equation}
where $c_N\in\bbr$ and
\[S_1 = S^{\,2} - \frac{2}{3}\openone .\]
Further, the structure of the recurrence relations hints to the following, rather natural splitting 
\begin{equation}
A_l = A^{\rm a}_l + A^{\rm d}_l\ , \qquad l=1,\ldots,N-1
\label{a_split}\end{equation}
of the matrix coefficients of the second Lax operator into $S$-commuting term $A^{\rm d}_l$ and some remainder
$A^{\rm a}_l$, see \cite{side9,yantih}. For the $S$-commuting term we have
\begin{equation}
A^{\rm d}_{l} = \left\{\begin{array}{rl}
a_{l} S_1,& l \equiv 0 \quad(\mod 2)\\ a_{l} S,& l\equiv 1 \quad (\mod 2) \,.
\end{array}\right.
\label{a_j_diag}\end{equation}
Above $a_{l}$, $l=1,\ldots,N-1$ are some scalar functions to be determined from the relations (\ref{rec_k}) and
(\ref{rec_1}).
\begin{rem}
Since $S(x,t)$ is a diagonalizable matrix (see Remark \ref{rem_constr}) we can deduce that the adjoint operator
$\ad_S$ is diagonalizable too --- its spectrum consists of $0,\pm 1, \pm 2$. This is why we can define the inverse
operator of $\ad_S$ on the $S$-noncommuting terms $A^{\rm a}_l$ as follows \cite{side9,yantih}:
\[\ad^{-1}_{S}A^{\rm a}_l = \frac{1}{4}\left(5\ad_{S}A^{\rm a}_l - \ad^3_{S}A^{\rm a}_l\right).\]
\end{rem}

Now, let us have a more detailed look of the equation (\ref{rec_k}). After substituting (\ref{a_split})
into (\ref{rec_k}) and taking into account that $(S_x)^{\rm d} =  (S_{1,x})^{\rm d} = 0$, we obtain
\begin{equation}
\left(\partial_xA^{\rm a}_j\right)^{\rm d} = - \left\{\begin{array}{rl}
\partial_xa_j S ,& j \equiv 1 \quad (\mod 2) \\ \partial_xa_j S_1 ,& j \equiv 0 \quad (\mod 2) 
\end{array}\right.\; ,\qquad j=2,\ldots,N-1
\label{rec_k_d}
\end{equation}
for the $S$-commuting part and
\begin{equation}
\left[S, A^{\rm a}_{j-1}\right] - \rmi\left(\partial_xA^{\rm a}_j\right)^{\rm a} =
\left\{\begin{array}{rl}
\rmi a_j S_x ,& j \equiv 1 \quad (\mod 2) \\ \rmi a_j S_{1,x} ,& j \equiv 0 \quad (\mod 2) \end{array}\right.
\label{rec_k_a}
\end{equation}
for the not commuting one. First, we solve (\ref{rec_k_d}) making use of the normalization relations
\[\tr S^2 = 2\, ,\qquad \tr S^2_1 = \frac{2}{3}\, \cdot	\]
The result for $a_j$ reads:
\begin{equation}
a_j = c_j - \left\{\begin{array}{rl}
\frac{1}{2}\,\partial^{-1}_x\tr\left[S(\partial_xA_j)^{\rm d}\right] ,& j \equiv 1 \quad (\mod 2) \\ \frac{3}{2}\,\partial^{-1}_x\tr\left[S_1(\partial_xA_j)^{\rm d}\right] ,& j \equiv 0 \quad (\mod 2) \end{array}\right.	
\label{a_k}\end{equation}
where the symbol $\partial^{-1}_x$ stands for any right inverse of the operator of partial differentiation in
variable $x$ and $c_j\in\bbr$ is an integration constant. After substituting (\ref{a_k}) into (\ref{rec_k_a}),
we obtain 
\begin{equation}
A^{\rm a}_{j-1} = \left\{\begin{array}{rl}
\Lambda A^{\rm a}_{j} + \rmi c_j\ad^{-1}_{S}S_{1,x}\, ,&
j\equiv 0 \quad (\mod 2)\\
\Lambda A^{\rm a}_{j} + \rmi c_j\ad^{-1}_{S}S_{x}\, ,&
j\equiv 1 \quad (\mod 2) 
\end{array}\right.
\label{A_km1}\end{equation}
where $\Lambda$ is an integro-differential operator defined as follows:
\begin{equation}
\Lambda \stackrel{\mathrm{def}}{=} \rmi\ad^{-1}_{S}\left\{ [\partial_x(.)]^{\rm a} - \frac{S_{x}}{2}
\partial^{-1}_x\tr\left[S(\partial_x(.))^{\rm d}\right] - \frac{3 S_{1,x}}{2}
\partial^{-1}_x\tr\left[S_1(\partial_x(.))^{\rm d}\right]\right\}.
\label{Lambda}\end{equation}
The $\Lambda$-operator introduced in (\ref{A_km1}) acts on the $S$-non commuting part of $A_j$ only but its
action can formally be extended on the $S$-commuting part as well by requiring 
\[\Lambda S \stackrel{\mathrm{def}}{=}  \rmi \ad^{-1}_{S} S_{x}\, ,\qquad
\Lambda S_{1} \stackrel{\mathrm{def}}{=}  \rmi \ad^{-1}_{S} S_{1,x}\,.\]
Taking into account (\ref{A_N}) and (\ref{A_km1}), one can verify that a NEE belonging to the integrable hierarchy
under consideration is generated through the equation
\begin{equation}
\rmi\ad^{-1}_{S} S_{t}	+ \sum_{j}c_{2j}\Lambda^{2j}S_1
+ \sum_{j}c_{2j-1}\Lambda^{2j-1} S = 0.
\label{int_hier}\end{equation}
The operator $\Lambda^2$ bear the name generating (recursion) operator of the integrable hierarchy. It can be
checked that (\ref{int_hier}) leads to system (\ref{ghf_nonloc1}) and (\ref{ghf_nonloc2}) after setting
$N=2$, $c_2 = -1$ and $c_1 = 0$. Thus, the system (\ref{ghf_nonloc1}) and (\ref{ghf_nonloc2}) is the simplest nontrivial
member of the family (\ref{int_hier}) indeed.

\section{Special Solutions}\label{solutions}

Here, we shall obtain some special solutions to the nonlocal equations (\ref{ghf_nonloc1}) and (\ref{ghf_nonloc2})
in explicit form. In doing this, we shall restrict ourselves with the class of trivial background solutions
satisfying the boundary condition:
\begin{equation}
\lim_{\vert x \vert\to \infty}u(x,t) = 0 \, ,\qquad \lim_{\vert x\vert\to\infty}v(x,t) = 1.
\label{triv_back}\end{equation}

Let us introduce the commuting operators
\begin{eqnarray}
L_0(\lambda) & = & \rmi\partial_x - \lambda S^{(0)}, \qquad 
S^{(0)}(x,t) = \left(\begin{array}{ccc}
0 & u_0(x,t) & v_0(x,t)\\ 
\varepsilon \overline{u_0(-x,t)} & 0 & 0 \\
\overline{v_0(-x,t)} & 0 & 0\end{array}\right), \label{lax1_bare}\\
A_0(\lambda) & = & \rmi\partial_t + \lambda A^{(0)}_1 + \lambda^2 A^{(0)}_2 ,
\qquad \lambda\in\bbc . \label{lax2_bare}
\end{eqnarray}
Above, $(u_0,v_0)$ is a pair of known dynamical fields solving the system (\ref{ghf_nonloc1}) and (\ref{ghf_nonloc2}) and fulfilling the additional constraint (\ref{constr}) and the boundary condition (\ref{triv_back}). Assume now that the $3\times 3$-matrix valued function
$\Psi_0$ ($\det \Psi_0(x,t,\lambda) = 1$) solves the problems 
\begin{eqnarray}
&& L_0(\lambda) \Psi_0(x,t,\lambda) =  0 \label{bare1}\\
&& A_0(\lambda)\Psi_0(x,t,\lambda) = \Psi_0(x,t,\lambda) f(\lambda)
\label{bare2}
\end{eqnarray}
to be called further in text bare problems. The polynomial 
\begin{equation}
f(\lambda) = - \frac{\lambda^2}{3} \diag(1,-2,1),
\label{disp_law}\end{equation}
appearing in the second bare problem, represents the dispersion law of the system (\ref{ghf_nonloc1}) and (\ref{ghf_nonloc2}). We refer the reader to \cite{yantih,yantih2} for more detailed explanations. 

Let us denote the set of all {\it bare} fundamental solutions by $\mathscr{F}_0$. For any $\Psi_0 \in\mathscr{F}_0$ we
construct new $3\times 3$-matrix valued function $\Psi_1 = \mathcal{G}\Psi_0$, where the unimodular $3\times 3$-matrix $\mathcal{G}(x,t,\lambda)$ is called dressing factor. The set of all such $3\times 3$-matrix valued functions will be
denoted by $\mathscr{F}_1$. In a natural way dressing transform induces an action on the Lax operators 
\begin{equation}
L_0 \to L_1 = \mathcal{G}L_0\mathcal{G}^{-1}\, ,\qquad A_0 \to A_1 = \mathcal{G}A_0\mathcal{G}^{-1}.
\label{lax_dres}\end{equation}
It is easily seen that $[L_1,A_1] = 0$ holds true. For dressing transform to be useful in constructing new solutions
to (\ref{ghf_nonloc1}) and (\ref{ghf_nonloc2}), we assume that any $\Psi_1\in\mathcal{F}_1$ is a fundamental solution to
\begin{equation}
\begin{split}
&L_1(\lambda) \Psi_1(x,t,\lambda) = 0\, , \\
&A_1(\lambda)\Psi_1(x,t,\lambda) = \Psi_1(x,t,\lambda)f(\lambda),
\end{split}	\label{dres}
\end{equation}
where the new Lax pair reads: 
\begin{eqnarray}
L_1(\lambda) & = & \rmi\partial_x - \lambda S^{(1)}, \qquad 
S^{(1)} = \left(\begin{array}{ccc}
0 & u_1(x,t) & v_1(x,t)\\ 
\varepsilon \overline{u_1(-x,t)} & 0 & 0 \\
\overline{v_1(-x,t)} & 0 & 0\end{array}\right), \label{lax1_dres}\\
A_1(\lambda) & = & \rmi\partial_t + \lambda A^{(1)}_1 + \lambda^2 A^{(1)}_2. \label{lax2_dres}
\end{eqnarray} 
The dynamical fields $u_1$ and $v_1$ above are some yet unknown solutions to (\ref{ghf_nonloc1}) and (\ref{ghf_nonloc2})
to be determined from the bare dynamical fields $u_0$ and $v_0$. 

It is not hard to convince ourselves that dressing factor fulfills the pair of partial differential equations: 
\begin{eqnarray}
\rmi\partial_x \mathcal{G} & - & \lambda \left(S^{(1)}\mathcal{G} - \mathcal{G}S^{(0)}\right) = 0\, ,
\label{g_pde1}\\
\rmi\partial_t \mathcal{G} & + & \sum_{k=1,2}\lambda^k \left( A^{(1)}_k \mathcal{G}
- \mathcal{G} A^{(0)}_k\right) = 0
\label{g_pde2}
\end{eqnarray}
that directly follows from (\ref{bare1}), (\ref{bare2}) and (\ref{dres}). Imposing certain natural requirements of regularity of dressing factor, see \cite{yantih2}, that system leads to the interrelation 
\begin{equation}
S^{(1)} = \mathcal{G}_{\infty}S^{(0)}\mathcal{G}^{-1}_{\infty},\qquad
\mathcal{G}_{\infty}(x,t) \stackrel{\mathrm{def}}{=} \lim_{\vert\lambda\vert\to\infty}\mathcal{G}(x,t,\lambda)
\label{s10_eq}\end{equation}
between the bare solution $S^{(0)}$ and the dressed one $S^{(1)}$, allowing one to determine $(u_1,v_1)$ from
$(u_0,v_0)$. 

Further on, we shall use a dressing factor that is a rational function of the form:
\begin{equation}
\mathcal{G}(x,t,\lambda) = \openone + \sum_{q}\left[\frac{\lambda B_q(x,t)}{\mu_q(\lambda - \mu_q)}
+ \frac{\lambda HB_q(x,t)H}{\mu_q(\lambda + \mu_q)}\right],
\qquad \mu^2_q\notin\bbr\, .
\label{dress_fac}
\end{equation}
That form is consistent with the symmetry relations:
\begin{eqnarray}
H\mathcal{G}(x,t,-\lambda)H &=& \mathcal{G}(x,t,\lambda)\,,\label{g_red1}\\
\mathcal{E}\mathcal{G}^{\dag}(-x,t,-\overline{\lambda})\mathcal{E} &=& \left[\mathcal{G}(x,t,\lambda)\right]^{-1}
\label{g_red2}\end{eqnarray}
which follow from the $\bbz_2\times\bbz_2$ reductions imposed on the bare and the dressed fundamental
solutions, see (\ref{psired1}) and (\ref{psired2}). 

The equation (\ref{s10_eq}) tells us that we can obtain $S^{(1)}$ if we know the residues of the dressing factor
(\ref{dress_fac}). In order to find $B_q(x,t)$, we first consider the identity
\[\mathcal{G}\mathcal{G}^{-1}=\openone . \]
Taking into account (\ref{g_red2}), the matrix inverse of (\ref{dress_fac}) is given by
\begin{equation}
\left[\mathcal{G}(x,t,\lambda)\right]^{-1} = \openone + \sum_p\left[\frac{ \lambda\mathcal{E}B^{\dag}_p(-x,t)\mathcal{E}}
{\overline{\mu}_p(\lambda + \overline{\mu}_p)} + \frac{\lambda\mathcal{E}HB^{\dag}_p(-x,t)H\mathcal{E}}
{\overline{\mu}_p(\lambda - \overline{\mu}_p)}\right].
\label{g_inv1}\end{equation}
Generally speaking $\mathcal{G}$ and $\mathcal{G}^{-1}$ have different simple poles.
After evaluating the left hand side of $\lim_{\lambda\to\overline{\mu}_p} (\lambda - \overline{\mu}_p)\mathcal{G}\mathcal{G}^{-1} = 0$, we derive the equations
\begin{equation}
\left[\openone + \sum_q\left(\frac{\overline{\mu}_p B_q(x,t)}{\mu_q(\overline{\mu}_p - \mu_q)}
+ \frac{\overline{\mu}_p HB_j(x,t)H}{\mu_q(\overline{\mu}_p + \mu_q)}\right)\right]\mathcal{E}HB^{\dag}_p(-x,t)H\mathcal{E} = 0	
\label{algsys1}\end{equation}
for the residues $B_p$. A similar evaluation of the residues at the poles $\pm\mu_p$ and $-\overline{\mu}_p$ does not lead to
essentially new equations this is why we shall ignore them. The equation (\ref{algsys1}) implies that the residues of $\mathcal{G}(x,t,\lambda)$ must be singular matrices, so the factorization
\begin{equation}
B_p(x,t)=X_p(x,t)F^T_p(x,t)
\label{b_fac}\end{equation}
holds for each of them. Above, $X_p(x,t)$ and $F_p(x,t)$ are two rectangular matrices and the superscript "$T$" stands for
matrix transposition. Using the factorization (\ref{b_fac}), we can reduce (\ref{algsys1}) to  
\begin{equation}
\mathcal{E}H\overline{F_p(-x)} = \overline{\mu}_p\sum_q\left(X_q(x)\frac{F^T_q(x)\mathcal{E}H\overline{F_p(-x)}}{\mu_q(\mu_q - \overline{\mu}_p)}
- HX_q(x)\frac{F^T_q(x)\mathcal{E}\overline{F_p(-x)}}{\mu_q(\overline{\mu}_p + \mu_q)}\right)
\label{matr_sys1}\end{equation}
which is viewed as a linear system for $X_q$. The linear system (\ref{matr_sys1}) is very easily solved when
(\ref{dress_fac}) has just two simple poles and $X$, $F$ are column-vectors. In that simplest case the result for $X$
is given by:
\begin{equation}
X(x) =	\left(\frac{\overline{\mu}F^T(x)\mathcal{E}H\overline{F(-x)}}{\mu(\mu - \overline{\mu})}
- \frac{\overline{\mu}F^T(x)\mathcal{E}\overline{F(-x)}}{\mu(\mu + \overline{\mu})}H\right)^{-1}\mathcal{E}H\overline{F(-x)} . 
\label{XF_sys}
\end{equation}

Next, we analyze the differential equation (\ref{g_pde1}) rewritten as
\begin{equation}
\lambda S^{(1)} = \rmi\partial_x \mathcal{G}\mathcal{G}^{-1} + \lambda\mathcal{G}S^{(0)}\mathcal{G}^{-1}.	
\end{equation}
It is not hard to prove \cite{yantih2} that the rectangular matrix $F_p(x,t)$ satisfies the rather simple
partial differential equation
\begin{equation}
\rmi\partial_x F^T_p = - \mu_p F^T_pS^{(0)}
\label{F_res}\end{equation}
which leads us to the conclusion that
\begin{equation}
F^T_p(x,t) = F^T_{p, 0}(t)\left[\Psi_0(x,t,\mu_p)\right]^{-1}\, .
\label{f_psi0}\end{equation}
for some matrix-valued functions $F_{p,0}$ of the variable $t$ only. It can be shown \cite{yantih2} that
the time evolution of $F_{p,0}$ is driven by the linear differential equation
\begin{equation}
\rmi\partial_t F^T_{p,0} = F^T_{p,0} f(\mu_p) 
\label{fj0_t}
\end{equation}
where $f(\lambda)$ is the dispersion law of (\ref{ghf_nonloc1}) and (\ref{ghf_nonloc2}), see (\ref{disp_law}).
It is immediately seen from (\ref{fj0_t}) that $F_{p,0}$ depends exponentially on time, therefore we just need
to make the following substitution
\begin{equation}
F^T_{p,0}\quad\to\quad F^T_{p,0}\rme^{-\rmi f(\mu_p)t}
\label{f_j0_evol}\end{equation}
to fully recover the time dependence in all equations containing the factors $F_{p,0}$.

In order to obtain explicit solutions of the system (\ref{ghf_nonloc1}) and (\ref{ghf_nonloc2}) we have to
pick up a convenient bare solution. In view of the boundary condition (\ref{triv_back}), a quite natural candidate
for a bare solution is the following one   
\begin{equation}
u_0(x,t) = 0,\qquad v_0(x,t) = 1 . 
\label{seed}
\end{equation}
For technical reasons, see \cite{yantih2}, we pick up
\begin{equation}
\Psi_0(x,t,\lambda) = \exp(- \rmi \lambda S_0 x) = \left(\begin{array}{ccc}
\cos{\lambda x} & 0 & - \rmi\sin{\lambda x} \\ 0 & 1 & 0 \\
- \rmi\sin{\lambda x} & 0 & \cos{\lambda x}
\end{array}\right)
\label{psi_0}\end{equation}
as a fundamental solution of the corresponding bare problems (\ref{bare1}) and (\ref{bare2}).
Further, we shall assume that (\ref{dress_fac}) has just two simple poles $\pm\mu$. Moreover, $X$ and $F$ will be column-vectors, so we can write down
\[X = \left(\begin{array}{c}
X^1 \\ X^2 \\ X^3
\end{array}\right),\qquad
F = \left(\begin{array}{c}
F^1 \\ F^2 \\ F^3
\end{array}\right)\]
for each of them. In view of (\ref{psi_0}) and (\ref{f_psi0}) $F$ acquires the form:  
\begin{equation}
F(x) = \left(\begin{array}{c}
F_{0}^{1}\cos\mu x + \rmi F_{0}^{3}\sin \mu x \\
F_{0}^{2} \\ F_{0}^{3}\cos \mu x + \rmi F_{0}^{1}\sin \mu x 
\end{array}\right),
\label{f_sol}
\end{equation}
where we have employed the notation $F^T_0= \left(F_{0}^{1}, F_{0}^{2}, F_{0}^{3}\right)$.

It turns out the form of the dressed solution depends on whether or not the second component of the vector
$F_{0}$ is zero. Let us consider first the case when $F_{0}^{2} = 0$ and $F_{0}^{1}\neq \pm F_{0}^{3}$. After recovering
the time dependence in (\ref{f_sol}) as in accordance with (\ref{f_j0_evol}), we see that $t$ appears in a common 
exponential factor only. So substituting (\ref{f_sol}) into (\ref{XF_sys}) and (\ref{s10_eq}), that factor cancels out
and we obtain the following stationary, nonsingular dressed solution:
\begin{eqnarray}
u_1(x) & = & 0\, , \label{u1_1}\\
v_1(x) & = & \left[\frac{c_1\cosh2\kappa x - c_2\sin2\omega x
+ \rmi (c_3\cos 2\omega x + c_4\sinh 2\kappa x)}{c_1\cosh2\kappa x
+ c_2\sin2\omega x - \rmi (c_3\cos 2\omega x - c_4\sinh 2\kappa x)}\right]^2 .
\label{v1_1}\end{eqnarray}
Above, we have introduced the additional notation
\begin{eqnarray}
\omega &=& \re\mu,\qquad \kappa=\im\mu,\qquad \varphi = \frac{\arg F_{0}^{1} - \arg F_{0}^{3}}{2} ,
\qquad \gamma = \frac{1}{2}\ln\vert F^1_{0}/F^3_{0}\vert,\label{auxpar1}\\ c_1 &=& \omega\sinh2\gamma,
\qquad c_2 = \kappa\cos2\varphi,\qquad
c_3 = \kappa\cosh2\gamma,
\qquad c_4 = \omega\sin2\varphi .
\label{c1234}\end{eqnarray}
Without loss of generality we can set $\omega >0$ and $\kappa >0$ in (\ref{auxpar1}).

It is not hard to see that when $F_{0}^{1} = \pm F_{0}^{3}$, i.e. when $\gamma = 0$ and $\varphi = 0, \pm \pi/2$,
we have that $c_1 = c_4 = 0$ while $c_2 = \pm \kappa$ and $c_3 = \kappa$. As a result, (\ref{u1_1}) and (\ref{v1_1})
degenerates to the bare solution (\ref{seed}), which is the reason why we required that $F_{0}^{1}\neq \pm F_{0}^{3}$.

Let us consider now the case when the second component of $F_{0}$ is not zero. After recovering the time evolution in
(\ref{f_sol}) and setting $F^2_{0} = 1$, we obtain 
\[F(x,t) = \left(\begin{array}{c}
\left(F_{0}^{1}\cos\mu x + \rmi F^3_{0}\sin\mu x\right)\exp\left(\frac{\rmi \mu^2 t}{3}\right) \\
\exp\left(-\frac{2\rmi \mu^2 t}{3}\right) \\ \left(F_{0}^3\cos\mu x + \rmi F_{0}^1\sin\mu x\right) 
\exp\left(\frac{\rmi \mu^2 t}{3}\right) 
\end{array}\right)\]
for the column-vector $F(x,t)$. The dressed solution now is given by:
\begin{eqnarray}
u_1 = \frac{\left[c_1\cosh2\kappa x - c_2\sin2\omega x
+ \rmi (c_3\cos 2\omega x + c_4\sinh 2\kappa x) - \varepsilon(\omega - \rmi \kappa)\rme^{2(2\omega\kappa t - \xi_0)}/2\right]}{\left[c_1\cosh2\kappa x
+ c_2\sin2\omega x - \rmi (c_3\cos 2\omega x - c_4\sinh 2\kappa x)
- \varepsilon(\omega + \rmi\kappa)\rme^{2(2\omega\kappa t - \xi_0)}/2\right]^2} \nonumber\\
\times \frac{2\rmi\omega\kappa}{\omega- \rmi \kappa}\rme^{-\rmi[(\omega^2 - \kappa^2)t + \delta_0]}
\left[\rme^{\rmi\varphi - \gamma}\cos(\omega + \rmi \kappa)x + \rmi \rme^{-\rmi\varphi + \gamma}
\sin(\omega + \rmi \kappa)x\right]\rme^{2\omega\kappa t - \xi_0}
\label{u1_2}\\
v_1 = \frac{\left[c_1\cosh2\kappa x - c_2\sin2\omega x
+ \rmi (c_3\cos 2\omega x + c_4\sinh 2\kappa x) - \varepsilon(\omega - \rmi \kappa)\rme^{2(2\omega\kappa t - \xi_0)}/2\right]}{\left[c_1\cosh2\kappa x
+ c_2\sin2\omega x - \rmi (c_3\cos 2\omega x - c_4\sinh 2\kappa x)
- \varepsilon(\omega + \rmi\kappa)\rme^{2(2\omega\kappa t - \xi_0)}/2\right]^2}\nonumber\\
\times \frac{\left\{\left[c_1\cosh2\kappa x - c_2\sin2\omega x
+ \rmi (c_3\cos 2\omega x + c_4\sinh 2\kappa x) - \varepsilon(\omega - \rmi \kappa)\rme^{2(2\omega\kappa t - \xi_0)}/2\right]\right.}
{\omega - \rmi\kappa} \\
\left.\times (\omega - \rmi \kappa) - 2\rmi\omega\kappa\varepsilon\rme^{2(2\omega\kappa t - \xi_0)}\right\}\nonumber
\label{v1_2}\end{eqnarray}
where  $\gamma$, $\varphi$ and the coefficients $c_j$, $j = 1,2,3,4$ are the same as in (\ref{auxpar1}) and (\ref{c1234}) respectively.  The two new parameters introduced above are defined through the equalities: 
\[\delta_0 = \frac{\arg F_{0}^{1} + \arg F_{0}^{3}}{2}, \qquad \xi_0 = \frac{1}{2}\ln\vert F_0^1F_0^3\vert \, .\]

\section{Conservation Laws}\label{consdens}

In this section we are going to view (\ref{ghf_nonloc1}) and (\ref{ghf_nonloc2}) as an infinite dimensional 
Hamiltonian system. More specifically, we are going to demonstrate how we can derive the conservation laws of
the nonlocal coupled system under consideration. For this to be done we apply a method proposed by Drinfel'd
and Sokolov \cite{DrSok*85}.

As already discussed in Remark \ref{rem_constr}, the matrix (\ref{lax1a}) has a simple spectrum --- its eigenvalues
are $0$, $\pm 1$. Thus, we can put $S(x,t)$ into a constant diagonal form by applying the gauge transformation:
\begin{eqnarray}
L(\lambda) &\to & \tilde{L}(\lambda) = \left[g(x,t)\right]^{-1}L(\lambda)g(x,t)
= \rmi \partial_x + \tilde{U}_0(x,t) - \lambda J,\label{l_tild}\\
A(\lambda) &\to & \tilde{A}(\lambda) = \left[g(x,t)\right]^{-1} A(\lambda)g(x,t) 
= \rmi\partial_t + \lambda\tilde{A}_1(x,t) + \lambda^2\tilde{A}_2(x,t),\label{m_tild}\\
g(x,t) &=& \frac{\sqrt{2}}{2}\left(\begin{array}{ccc}
1 & 0 & -1 \\ \varepsilon \overline{u(-x,t)} & \sqrt{2} v(x,t) & \varepsilon \overline{u(-x,t)}\\
\overline{v(-x,t)} & -\sqrt{2} u(x,t)  & \overline{v(-x,t)}
\end{array}\right),\quad J  = \left(\begin{array}{ccc}
1 & 0 & 0 \\ 0 & 0 & 0 \\ 0 & 0 & -1
\end{array}\right)
\label{g_trans}
\end{eqnarray}
on the Lax operators (\ref{lax_gen1}) and (\ref{lax2a}). The explicit form of $\tilde{U}_0$ to be used further
is given by
\begin{equation}
\tilde{U}_0(x,t)  = \frac{\rmi}{2}\left(\begin{array}{ccc}
\mathcal{B}(x,t) & \mathcal{C}(x,t) & B(x,t) \\
-\varepsilon \overline{\mathcal{C}(-x,t)} & -2\mathcal{B}(x,t) 
& -\varepsilon \overline{\mathcal{C}(-x,t)}\\
\mathcal{B}(x,t) & \mathcal{C}(x,t) & \mathcal{B}(x,t)
\end{array}\right)
\label{u_0}\end{equation}
where $\mathcal{C}(x,t) = \sqrt{2}[u(x,t)v_x(x,t) - v(x,t)u_x(x,t)]$ and $\mathcal{B}(x,t)$ is the same as in
(\ref{cal_b}).

Let us apply Drinfel'd and Sokolov's method of diagonalization of Lax pair to derive the integrals of motion
of (\ref{ghf_nonloc1}) and (\ref{ghf_nonloc2}). For that purpose we introduce the gauge transform
\begin{equation}
\mathcal{P}(x,t,\lambda)= \openone +\frac{P_1(x,t)}{\lambda} +
\frac{P_2(x,t)}{\lambda^2} + \cdots \label{P_trans}
\end{equation}
All the coefficients $P_l$ ($l=1,2,\ldots$) appearing above are some ff-diagonal $3\times 3$ matrices to be
determined further. As a result of the action of $\mathcal{P}$, we have 
\begin{eqnarray}
\mathcal{L} & = & \hat{\mathcal{P}}\tilde{L}\mathcal{P}
= \rmi \partial_x - \lambda J + \mathcal{L}_0
+ \frac{\mathcal{L}_1}{\lambda} + \cdots,\label{L_1}\\
\mathcal{A} & = & \hat{\mathcal{P}}\tilde{A}\mathcal{P}
= \rmi \partial_t + \lambda^{2} \mathcal{A}_{-2} + \lambda \mathcal{A}_{-1}
+ \mathcal{A}_{0} + \frac{\mathcal{A}_{1}}{\lambda} + \frac{\mathcal{A}_{2}}{\lambda^2} + \cdots
\label{Mu}
\end{eqnarray}
All the coefficients $\mathcal{L}_k$, $\mathcal{A}_{k}$, $k=-1,0,1,\ldots$ are required to be
diagonal matrices. Then the compatibility condition $[\mathcal{L}, \mathcal{A}] = 0$
is equivalent to the equations
\[\partial_t\mathcal{L}_k - \partial_x \mathcal{A}_k =  0,\qquad
k=0,1,\ldots.\]
These equations mean that $\mathcal{L}_k$ are conserved densities of the system (\ref{ghf_nonloc1}) and
(\ref{ghf_nonloc2}) while $\mathcal{A}_k$ are the corresponding currents.

Let us now rewrite the equality (\ref{L_1}) in the following way:
\begin{equation}
\tilde{L}\mathcal{P} = \mathcal{P}\mathcal{L}
\label{lppl}
\end{equation}
Since (\ref{lppl}) should hold identically in $\lambda$, it splits into the following recurrence relations:
\begin{eqnarray}
\lambda^{0}  &:& \tilde{U}_0 - J P_1 = \mathcal{L}_0 - P_1 J,\\
\lambda^{-1} &:& \rmi P_{1,x} + \tilde{U}_0 P_1 - JP_2
= \mathcal{L}_1 + P_1\mathcal{L}_0 - P_2 J, \label{lambda_m1}\\
& & \ldots \nonumber\\
\lambda^{-k} &:& \rmi P_{k,x} + \tilde{U}_0 P_k - JP_{k+1}
= \mathcal{L}_k - P_{k+1}J + \sum^{k-1}_{m = 0}P_{k-m}\mathcal{L}_m,
\label{lambda_mk}\\ 
&&\cdots \nonumber
\end{eqnarray}
In order to resolve those relations, one should split each relation into a diagonal and
off-diagonal part. For example, from the first relation above one has
\begin{equation}
\mathcal{L}_0 = \tilde{U}^{\rm d}_0,\qquad \tilde{U}^{\rm a}_0 =  [J,P_1]
\label{recursplit_1}
\end{equation}
where the superscripts $\rm d$ and $\rm a$ above denote projection onto diagonal and
off-diagonal part of a~matrix respectively. Taking into account the explicit
form of $\tilde{U}_0$ (formula (\ref{u_0})) for $\mathcal{L}_0$ we have
\[\mathcal{L}_0 = \frac{\rmi}{2}\mathcal{B}(x,t)\left(\begin{array}{ccc}
1 & 0  & 0 \\
0 & -2 & 0 \\
0 & 0  & 1
\end{array}\right).\]
Thus, as a density $\mathcal{I}_1$ of the first integral of motion we can choose the quantity $- \mathcal{B}(x,t)$. 

On the other hand, after inverting the commutator in the second equation in (\ref{recursplit_1}), one obtains
\begin{equation}
P_1 = \frac{\rmi}{2}\left(\begin{array}{ccc}
0  & \mathcal{C}(x,t) & \mathcal{B}(x,t)/2 \\
\varepsilon \overline{\mathcal{C}(-x,t)} & 0 & -\varepsilon \overline{\mathcal{C}(-x,t)}\\
-\mathcal{B}(x,t)/2 & -\mathcal{C}(x,t) & 0
\end{array}\right).
\label{t_1}
\end{equation}
Similarly, for $\mathcal{L}_1$ one needs to extract the diagonal part of (\ref{lambda_m1}). The result reads
\begin{equation}
\mathcal{L}_1 = \left(\tilde{U}^{\rm a}_0 P_1\right)^{\rm d}.
\label{recursplit_2d}
\end{equation}
After substituting the expression (\ref{t_1}) for $P_1$ into (\ref{recursplit_2d}), one obtains
\[\mathcal{L}_1 = \frac{1}{8}\left[\mathcal{B}^2(x,t)
- 4\varepsilon \mathcal{C}(x,t)\overline{\mathcal{C}(-x,t)}\right]
\left(\begin{array}{ccc}
1 & 0 & 0 \\
0 & 0 & 0 \\
0 & 0 & -1
\end{array}\right).\]
Therefore one can pick up
\[\mathcal{I}_2 = \mathcal{B}^2(x,t) - 4\varepsilon \mathcal{C}(x,t)\overline{\mathcal{C}(-x,t)}\]
as a second conserved density. Proceeding further, one is able to find a conserved density of arbitrary order.
It is seen from (\ref{lambda_mk}) that $k$-th coefficient $\mathcal{L}_k$ is derived from the relation
\begin{equation}
\mathcal{L}_k = \big(\tilde{U}^{\rm a}_0P_k\big)^{\rm d}
\label{l_k}\end{equation}
while $P_k$ is obtained from the recurrence relation:
\[P_k = \ad^{-1}_{J}\left(\rmi \partial_x P_{k-1}
+ (\tilde{U}_0 P_{k-1})^{\rm a} - \sum^{k-2}_{l=0}P_{k-1-l}\mathcal{L}_l\right).\]

In order to obtain the integrals of motion of the system (\ref{ghf_nonloc1}) and (\ref{ghf_nonloc2}), one just
needs to integrate the conserved densities over the variable $x$ as given below:
\[I_k = \int^{\infty}_{-\infty}\rmd x \,\mathcal{I}_k(x,t), \qquad k=1,2,\ldots\]
The limits of integration implies certain choice of boundary conditions imposed on the dynamical
fields $u$ and $v$. Above, we have had in mind the trivial background condition as introduced in (\ref{triv_back}).

\section{Conclusion}

In the present preprint we have introduced and studied the nonlocal reduction (\ref{ghf_nonloc1}) and
(\ref{ghf_nonloc2}) of a generalized Heisenberg ferromagnet equation. The integrable hierarchy
corresponding to (\ref{ghf_nonloc1}) and (\ref{ghf_nonloc2}) has been completely described in terms of
generating operators, see (\ref{int_hier}). Furthermore, we have constructed special solutions to
(\ref{ghf_nonloc1}) and (\ref{ghf_nonloc2}) by applying dressing procedure. We have derived only special
solutions related to 4 complex discrete eigenvalues of the scattering operator. However, there exist
two more types of solutions: doublet solutions and quasi-rational solutions that are related to pairs
of discrete eigenvalues. Constructing those types of solutions will be demonstrated elsewhere. By similar
fashion, one can derive special solutions to any member of the integrable hierarchy --- the only difference
will be the time dependence of the solutions. 

Another important issue concerns the Hamiltonian formalism and the integrals of motion for the system of
nonlocal equations under consideration. By using Drinfel'd-Sokolov's method of Lax pair diagonalization,
recursive formulas to derive the conserved densities of (\ref{ghf_nonloc1}) and (\ref{ghf_nonloc2}) have
been obtained, see (\ref{l_k}). Generally speaking those conserved densities contain nonlocal terms which
correspond to the nonlocal terms in the NEEs under consideration.

\section*{Acknowledgments}
The work has been supported by grant DN 02--5 of Bulgarian Fund "Scientific Research".


\begin{thebibliography}{99}

\bibitem{ablomus}
Ablowitz, M. and Musslimani, Z., Integrable Nonlocal Nonlinear Schr\"odinger Equation, {\it Phys. Rev. Lett.}
{\bf 110} (2013) 064105(5).
\bibitem{BoPo90}
Borovik, A. E. and Popkov, V. Yu., Completely Integrable Spin - 1 Chains, {\it Sov. Phys. JETPH} {\bf 71}
(1990) 177--85.
\bibitem{DrSok*85}
Drinfel'd, V. G. and Sokolov, V. V., Lie Algebras and Equations of Korteweg-de Vries Type,
{\it Jour. Math. Sci.} {\bf 30} (1985) 1975--2036.
\bibitem{fok}
Fokas, A., Integrable Multidimensional Versions of the Nonlocal Nonlinear Schr\"odinger Equation,
{\it Nonlinearity} {\bf 29}, 2 (2016) 319--324.
\bibitem{book}
Gerdjikov, V., Vilasi, G. and Yanovski, A., \emph{Integrable Hamiltonian Hierarchies. Spectral and
Geometric Methods}, Lecture Notes in Physics {\bf 748}, Springer, Berlin, 2008.
\bibitem{ggi}
Gerdjikov, V. S., Grahovski, G. G. and Ivanov, R. I., The N-wave Equations with PT Symmetry,
{\it Theor. Math. Phys.} {\bf 188}, 3 (2016) 1305--1321.
\bibitem{side9}
Gerdjikov, V., Grahovski, G., Mikhailov, A. and Valchev, T., Polynomial Bundles and Generalized Fourier
Transforms for Integrable Equations on {\bf A.III}-type Symmetric Spaces, {\it SIGMA} {\bf 7}, 096 (2011) 48 pages.
\bibitem{gurpec}
G\"urses, M. and Pekcan, A., Integrable Nonlocal Reductions, in: {\it Symmetries, Differential Equations
and Applications}, Springer Proc. in Mathematics \& Statistics {\bf 266}, Eds.: V. Kac, P. Olver,
P. Winternitz and T. \"Ozer, Springer, Cham, 2018, 27--52.
\bibitem{ishim}
Ishimori, Y., Multi-vortex Solutions of a Two-dimensional Nonlinear Wave Equation, {\it Prog. Theor. Phys.}
{\bf 72} (1984) 33-–37.
\bibitem{mikh2}
Mikhailov, A. V., The Reduction Problem and Inverse Scattering Method, {\it Physica D} {\bf 3} (1981) 73--117.
\bibitem{myr1}
Myrzakulov, R., Vijayalarshmi, S., Nugmanova,  G. and Lakshmanan, M., A (2+1) Dimensional Integrable
Spin Models with Self-consistent Potentials, {\it Symmetry} {\bf 7} (2015) 1352--1373.
\bibitem{myr2}
Myrzakulov, R., Mamyrbekova, G., Nugmanova, G. and Lakshmanan, M., Integrable (2+1) Dimensional 
Spin Model: Geometric and Gauge Equivalent Counterparts, Solitons and Coherent Structures, {\it Phys. Lett. A}
{\bf 233} (1997) 391.
\bibitem{blue-bible}
Takhtadjan, L. and Faddeev, L., \emph{The Hamiltonian Approach to Soliton Theory}, Springer Verlag, Berlin, 1987.
\bibitem{yan}
Yanovski, A. B., On the Recursion Operators for the Gerdjikov, Mikhailov and Valchev System, {\it Jour. Math. Phys.}
{\bf 52}, 8 (2011) 082703.
\bibitem{yantih}
Yanovski, A. B. and Valchev, T. I., Pseudo-Hermitian Reduction of a Generalized Heisenberg Ferromagnet Equation.
I. Auxiliary System and Fundamental Properties, {\it Jour. Nonl. Math. Phys.} {\bf 25} (2018) 324--350 , arXiv: 1709.09266 [nlin.SI].
\bibitem{tih}
Valchev, T., On Mikhailov's Reduction Group, {\it Phys. Lett. A} {\bf 379} (2015) 1877-–1880.
\bibitem{yantih2}
Valchev, T. and Yanovski, A., Pseudo-Hermitian Reduction of a Generalized Heisenberg Ferromagnet Equation.
II. Special Solutions, {\it Jour. Nonl. Math. Phys.} {\bf 25} (2018) 442--461, arXiv 1711.06353v2.
\bibitem{yanvil}
Yanovski, A. and Vilasi, G., Geometry of the Recursion Operators for the GMV System, {\it Jour. Nonl. Math. Phys.}
{\bf 19}, 3 (2012) 1250023-1/18. 
\bibitem{zakh-mikh}
Zakharov, V. and Mikhailov, A., On the Integrability of Classical Spinor Models in Two-Dimensional Space-Time,
{\it Commun. Math. Phys.} {\bf 74} (1980) 21--40.
\bibitem{ZS}
Zakharov, V. E. and Shabat, A. B., A Scheme for Integrating Nonlinear Equations of Mathematical Physics by
the Method of the Inverse Scattering Transform II, {\it Funct. Anal. and Appl.} {\bf 13} (1979) 13--23.

\end{thebibliography}
\end{document}